\documentclass[letterpaper, 10 pt, conference]{ieeeconf}  % Comment this line out if you need a4paper

\IEEEoverridecommandlockouts                              % This command is only needed if 
\usepackage{graphics} % for pdf, bitmapped graphics files
\usepackage{epsfig} % for postscript graphics files
\usepackage{mathptmx} % assumes new font selection scheme installed
\usepackage{times} % assumes new font selection scheme installed
\usepackage{amsmath} % assumes amsmath package installed
\usepackage{amssymb}  % assumes amsmath package installed
\usepackage{url}

\title{\LARGE \bf
Functional Connectivity Dynamics show Resting-State Instability and Rightward Parietal Dysfunction in ADHD 
}

\author{Rohit Misra$^{1}$ and Tapan K. Gandhi$^{1}$% <-this % stops a space
\thanks{$^{1}$Department of Electrical Engineering, Indian Institute of Technology Delhi, Hauz Khas, New Delhi, India, 110016}%
}

\begin{document}
\bstctlcite{IEEEexample:BSTcontrol}

\maketitle
\thispagestyle{empty}
\pagestyle{empty}

%%%%%%%%%%%%%%%%%%%%%%%%%%%%%%%%%%%%%%%%%%%%%%%%%%%%%%%%%%%%%%%%%%%%
\begin{abstract}
Attention Deficit/Hyperactivity Disorder (ADHD) is one of the most common neurodevelopmental disorders in children and is characterised by inattention, impulsiveness and hyperactivity. While several studies have analysed the static functional connectivity in the resting-state functional MRI (rs-fMRI) of ADHD patients, detailed investigations are required to characterize the connectivity dynamics in the brain. In an attempt to establish a link between attention instability  and the dynamic properties of Functional Connectivity (FC), we investigated the differences in temporal variability of FC between 40 children with ADHD and 40 Typically Developing (TD) children. Using a sliding-window method to segment the rs-fMRI scans in time, we employed seed-to-voxel correlation analysis for each window to obtain time-evolving seed connectivity maps for seeds placed in the posterior cingulate cortex (PCC) and the medial prefrontal cortex (mPFC). For each subject, the standard deviation of the voxel connectivity time series was used as a measure of the temporal variability of FC. Results showed that ADHD patients exhibited significantly higher variability in dFC than TD children in the cingulo-temporal, cingulo-parietal, fronto-temporal, and fronto-parietal networks ($p_{FWE} < 0.05$). Atypical temporal variability was observed in the left and right temporal gyri, the anterior cingulate cortex, and lateral regions of the right parietal cortex. The observations are consistent with visual attention issues, executive control deficit, and rightward parietal dysfunction reported in ADHD, respectively. These results help in understanding the disorder with a fresh perspective linking behavioural inattention with instability in FC in the brain.
% \newline
% \indent \textit{Clinical relevance}— .
\end{abstract}

\section{INTRODUCTION}
Attention Deficit Hyperactivity Disorder (ADHD) is characterized by age-inappropriate inattention, impulsiveness and hyperactivity in both children and adults. According to recent studies \cite{nigg2001adhd}, ADHD is among the most common neurodevelopmental disorders and affects around 5-11\% of children. The disorder is commonly associated with an inability to sustain attention for a long duration which impacts their professional, social, and personal lives \cite{nigg2001adhd}. Advances in neuroimaging techniques have enabled researchers to investigate the neural correlates of the behavioral patterns observed in ADHD. Neural activity deficits are consistently reported in the fronto-striatal and fronto-parietal circuits in ADHD patients, along with abnormalities in the medial frontal, anterior cingulate cortex, and inferior parietal regions \cite{Dickstein2006, Dillo2010}. Electrophysiological investigations have also reported reduced intensities in error positivity Event Related Potentials (ERPs) in ADHD patients \cite{herrmann2010neural}. 

In addition to behavioral, task-based investigations, neuroimaging analyses of the brain in the resting-state also provide an opportunity to study the neurological bases of the unrest and hyperactivity reported in ADHD patients. Resting-State Functional Connectivity (RSFC) \cite{Biswal1995} analysis has proved to be effective in understanding the spontaneous widespread activity in the brain \cite{vanden}. To study the Functional Connectivity (FC), the temporal synchronization between Blood Oxygen Level Dependent (BOLD) time series for different brain regions is considered as a marker for functional pairing between the regions. However, determining the correlations over the BOLD series of entire scans of 5-10 minutes leads to loss of information regarding the dynamic evolution of brain networks during the scan.
The study of the dynamic properties of functional connectivity is a very recent development in the field of neuroimaging and has been since used extensively in highlighting the evolving nature of functional connectivity in the brain. Dynamic functional connectivity (dFC) studies have have shown significant differences in the dynamic properties exhibited by patients of schizophrenia \cite{damaraju2014dynamic}, Alzheimer's disorder \cite{Gu2020}, autism spectrum disorder \cite{Li2020}, and various other neuropsychiatric disorders \cite{Greicius2008}.

% The study of dFC is commonly done by distributing an entire resting state or task-based fMRI scan into sequentially cascaded windows. FC is determined for each of these windows using a metric of connectivity such as the Pearson correlation coefficient, precision matrices, or mutual information. The temporal evolution of window-wise connectivity is used to extract dynamic features of brain function. Some studies have also employed seed-to-voxel-based connectivity maps to study dFC \cite{kaiser2016dynamic} instead of brain parcellation methods that generate connectivity matrices. In addition to the above-mentioned approaches to dFC, there have been several attempts to extract FC dynamics using point-process models (\cite{lindquist2014evaluating}), dynamical system models (\cite{Power}, \cite{Jia}, \cite{NIELSEN2018116}), and Bayesian models (\cite{Andersen}, \cite{Ryan}).  

Investigations into the neural activity that differentiates patients with ADHD from Typically Developing (TD) children have revealed several abnormalities in functional connectivity and its dynamics. ADHD patients have been reported to show atypical resting state connectivity in the medial regions of the default mode network (DMN), insular cortex, and the dorsolateral prefrontal cortex (dlPFC) \cite{Fair2013}. In addition to the DMN, insufficient suppression of inter-network connectivity between the DMN and the dlPFC has also been observed in ADHD patients \cite{Hoekzema2014}. Atypical neural activity in the dorsal attention network, reward-motivation circuits, and the executive control network (anterior cingulate gyrus) \cite{Mostert2016} has also been linked to ADHD patients.

Multiple attempts to understand the temporal dynamics of functional connectivity have resulted in further insights into the neural bases of ADHD. Significant differences in static and dynamic FC have been noted in the inferior, middle, and superior temporal gyri, medial frontal gyri, insula, anterior and posterior cingulate cortices (ACC and PCC) \cite{Ahmadi2021}. Atypical dynamics in the cingulo-opercular and the sensorimotor networks have also been reported along with high dFC fluctuations in the ventral medial PFC and the anterior medial PFC \cite{Sun2021}. Overall, ADHD patients show weaker correlations in the DMN, attention, and visual networks, which are characterized by increased functional integration and reduced segregation on a global network scale in the brain \cite{Zhu2023}. Besides investigatory studies, Wang and colleagues have used biomarkers extracted from dFC to propose a diagnostic model for ADHD \cite{Wang2018}.

In order to examine the dynamics of resting-state FC in ADHD patients further, in this study, we attempt to establish a link between the behavioral instability, inattentiveness, and hyperactivity in ADHD and the dynamic properties of the default mode network in the resting state. We hypothesize that the functional connectivity dynamics of ADHD patients will show higher temporal variability as compared to typically developing (TD) children. We use a seed-to-voxel correlation-based connectivity analysis, aided by a sliding window dFC paradigm, to extract the brain-wide time-evolving connectivity with the seed. Here we report the results of two analyses with seeds placed at the prominent nodes of the DMN, namely the posterior cingulate cortex (PCC) and the medial prefrontal cortex (mPFC).

\section{MATERIALS AND METHODS}
\begin{figure*}[h!]
    \centering
    \includegraphics[width = 0.9\textwidth]{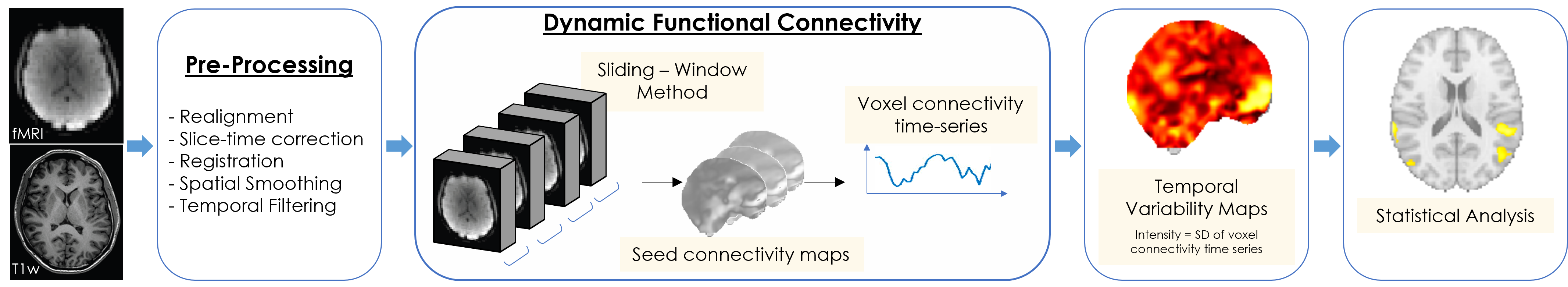}
    \caption{Steps performed to extract temporal variability in functional connectivity (FC) and subsequent statistical comparison of ADHD and TD groups. After pre-processing, the sliding window method is used to obtain seed connectivity maps. The standard deviation (SD) of Fisher's z-transformed voxel connectivity time series is used to quantify temporal variability in FC.}
    \label{pipeline}
\end{figure*}

\begin{figure}[h!]
    \centering
    \includegraphics[width = 0.8\columnwidth]{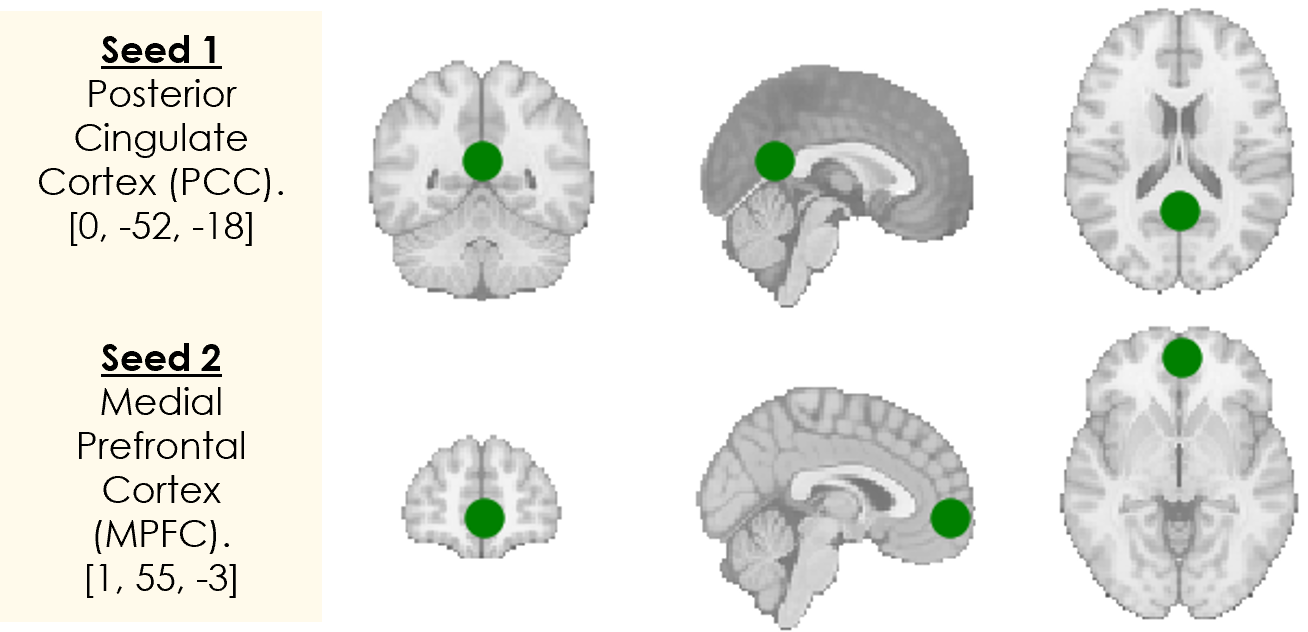}
    \caption{Spherical ROIs used in this study for seed-to-voxel functional connectivity analysis. (a) Posterior Cingulate Cortex (PCC): Centre at [0,-52,-18], with radius = 8mm, and (b) Medial PreFrontal Cortex (mPFC): Centre at [1, 55, -3], with radius = 8mm.}
    \label{seeds}
\end{figure}

\subsection{Magnetic Resonance Imaging Data}
In this study, structural and functional MRI scans of 40 children with ADHD-Combined ($11.47\pm 2.86$ years, 5  Females) and 40 typically developing ($13.10\pm 3.11$ years, 21 Females) children were used from the ADHD-200 dataset \cite{adhd}. The scans were acquired using a Seimens Allegra MRI scanner. The anatomical volumes were T1 weighted acquired with a magnetization prepared rapid-acquisition gradient-echo sequence with a resolution of $1.3 \times 1.3 \times 1.3 \ mm^3$. The resting-state functional Magnetic Resonance Imaging (rs-fMRI) scans were T2* weighted (BOLD weighted) multi-echo scans acquired for a duration of approximately 6 minutes with a repetition time (TR) = 2 seconds and resolution $3 \times 3 \times 3 \ mm^3$. For each rs-fMRI scan, we analysed the first 176 volumes to maintain data-volume consistency across subjects.
\subsection{Pre-Processing}
The fMRI data were pre-processed to remove artefacts, noise, and potential confounds from further analysis. Using the SPM12 MATLAB toolbox \cite{spm}, the scans were first corrected for head motion and the volumes were realigned. Slice-time correction was performed, followed by co-registration of the scans with the respective anatomical scans. The co-registered fMRI volumes were normalised by warping them to the standard MNI template. Further, spatial smoothing was done using an isotropic Gaussian kernel with full-width half maximum (FWHM) = 8mm to reduce the spatial noise. 

Temporal pre-processing was done using custom Python scripts. The BOLD time series were linearly detrended to reduce the effect of noise and drift followed by temporal band-pass filtering (0.01 - 0.1 Hz) using a Butterworth filter. Finally, the time series were standardized to their z-score values.
\subsection{Dynamic Functional Connectivity}
The functional connectivity in the brain was analysed using seed-to-voxel-based correlation. A seed Region of Interest (ROI) was identified using a spherical mask of radius 8mm surrounding the seed coordinates. The ROI time series was extracted by determining the average time series of the voxels present in the seed ROI. In this study, the first seed was selected as the PCC, the hub of the DMN, with MNI-space coordinates [0, -52, -18]. For the second analysis, the seed was selected as the mPFC ([0, -52, -18]), which has also been repeatedly reported to show atypical connectivity in ADHD patients \cite{Ahmadi2021, Sun2021} (Figure \ref{seeds}). 

To extract the dynamic FC, the ROI time series and the voxel-wise time series were segmented using a sliding window method. The tapered-rectangular window of duration 100s (50 TRs) was generated by convolving a rectangular window of duration 80s (40 TRs) with a Gaussian kernel of FWHM = 20s (10 TRs). The step-size of the sliding window was selected as 2s (1 TR) to consequently generate 127 windows from the rs-fMRI scans of 176 volumes. For each window, the Pearson correlation of the windowed ROI time series with the windowed voxel time series was calculated to generate a seed connectivity map. Therefore, for each scan, the temporal evolution of seed connectivity maps provided the dFC of all regions in the brain with the seed. Each seed connectivity map was Fisher's z-transformed to stabilize the variance of the correlation values. Next, utilizing the standard deviation of the transformed voxel connectivity time series as a measure of temporal variability of voxel-wise functional connectivity with the seed, a map of seed-to-voxel FC temporal variability was obtained for each subject. The processing pipeline is depicted in Figure \ref{pipeline}.

\subsection{Statistical Analysis}
Statistical analyses were performed using the SPM12 toolbox. Voxel-wise comparison of the temporal variability maps of the two groups ADHD and TD was made using a one-tailed unpaired two-sample t-test. Using the contrast maps of temporal variability for ADHD $>$ TD, the significant clusters were identified and the effect of interest was obtained by thresholding the t-score maps at the Family-Wise Error (FWE) corrected threshold. Age and sex of the subjects were added as covariates of no interest in the tests to eliminate confounding effects. Cluster-based thresholding was applied at $p_{unc} < 0.01$ and $p_{unc} < 0.05$ for the seeds placed at PCC and mPFC, respectively, followed by FWE correction at $p_{FWE} < 0.05$ for both seeds. Consequently, the cluster extent threshold obtained was used to generate the corrected statistical maps with significant clusters.

\section{RESULTS}

Group wise comparison of temporal variability of FC among the ADHD and TD cohorts results in three major clusters where ADHD subjects showed significantly elevated temporal variability in dFC (see Figure (CITATION FIGURE)). With the seed placed at PCC, we found two significant clusters with increased dynamics in FC in ADHD with a cluster extend threshold of 6580 voxels. The first cluster exhibited a peak t-score of 4.22 in the left inferior temporal gyrus ([-57, -64, 2]) and expanded dorsally to encapsulate the left middle and superior temporal gyri along with the left parietal operculum. The second cluster had a peak t-score of 3.56 (at [39, -44, -12]) and encompassed the right inferior temporal gyrus, parietal operculum, supramarginal gyrus, and lateral and superior parietal regions. When the mPFC was selected as the seed, one significant cluster was obtained with a cluster extent threshold of 25,162 voxels. With a peak t-score of 3.44 at [-3, 18, -6], the widespread cluster comprised the anterior cingulate cortex (ACC), the anterior subcallosal gyrus, the right medial temporal gyrus, and regions in the lateral parietal cortex. There were no significant clusters that showed lower variability in the ADHD cohort compared to TD children for either of the seeds.
\begin{figure}[!h]
    \centering
    \includegraphics[width = 0.8\columnwidth]{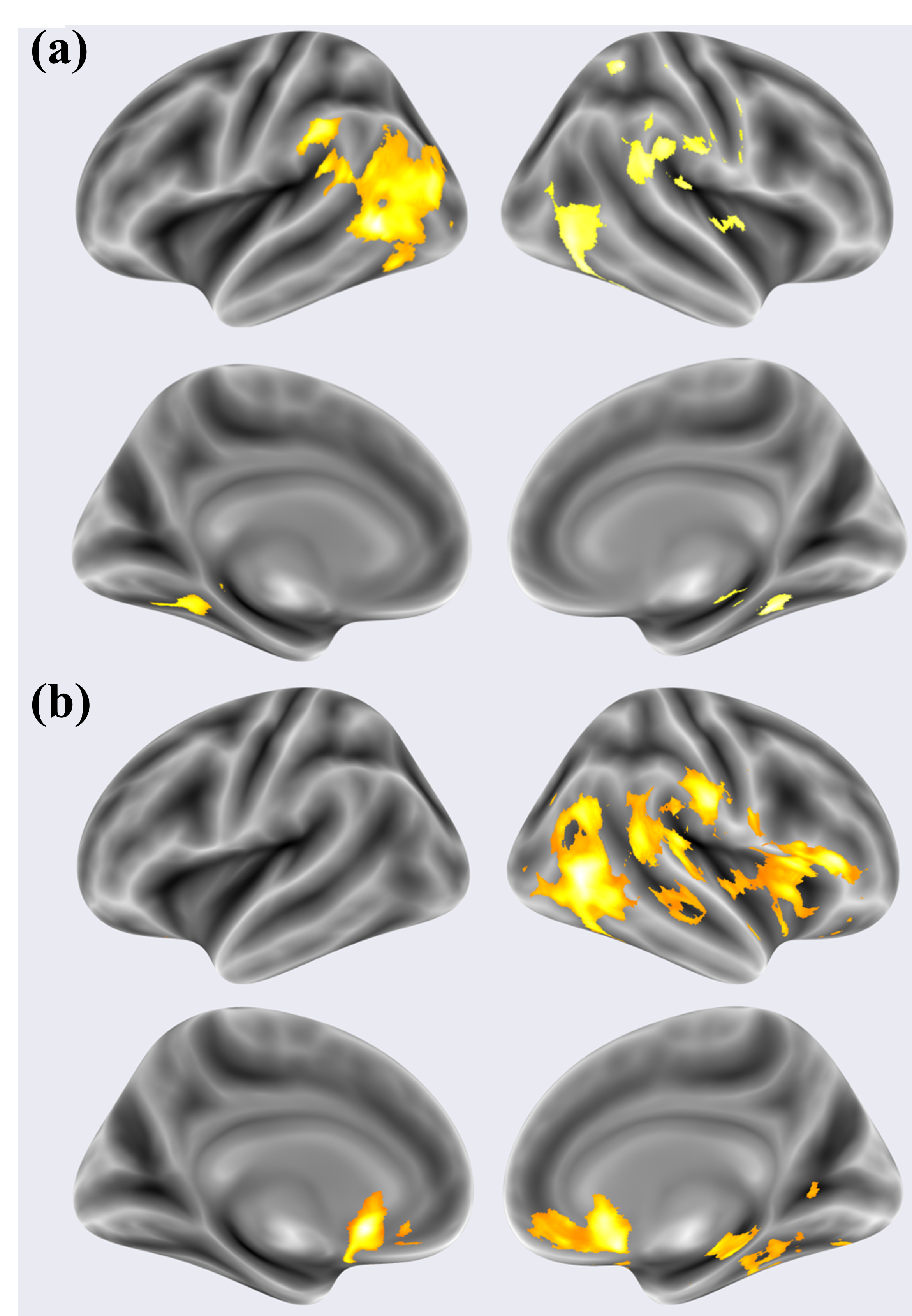}
    \caption{Inflated brain statistical maps showing significant clusters for contrast ADHD $>$ TD. \textbf{(a)} Regions with high variability in connectivity with PCC in ADHD subjects ($p_{unc} < 0.01$, $p_{FWE} < 0.05$) \textbf{(b)} Regions with high variability in connectivity with mPFC in ADHD subjects ($p_{unc} < 0.05$, $p_{FWE} < 0.05$). (\textbf{Key}: Top Left = Left Hemisphere Lateral View, Top Right = Right Hemisphere Lateral View, Bottom Left = Left Hemisphere Medial View, Bottom Right = Right Hemisphere Medial View) }
    \label{results}
\end{figure}

\section{DISCUSSION}
By means of this study, we aimed to investigate the link between the behavioural inattentiveness and hyperactivity associated with ADHD patients and the dynamic fluctuations in the functional connectivity in their brains. The seeds were chosen based on their prominence in the DMN, which is implicated in introspective and mind-wandering states of the brain. The results obtained highlight that ADHD patients exhibit high temporal variability in cingulo-temporal, cingulo-parietal, fronto-temporal, and fronto-parietal functional connectivity. High FC fluctuations in these regions signify instability within the DMN and the fronto-parietal attention networks. The results reinforce the findings using static FC that report atypical connectivity in the DMN \cite{Fair2013} and abnormal inter-network FC \cite{Zhu2023} in ADHD patients. 

We also found bilateral increase in cingulo-temporal FC dynamics in the ADHD group. The inferior and middle temporal gyri are generally involved in multi-modal processing, including familiar facial recognition, visual-attention, emotion and social processing, and semantic memory \cite{Ahmadi2021}. Abnormal activity in the temporal gyri has also been suggested as a possible compensatory mechanism in ADHD patients \cite{wang2013dysfunctional}. ADHD patients also showed more temporal instability in the connectivity between the anterior regions of the mPFC and the ACC. Given the involvement of ACC in the executive control network and focused task-directed behaviour, such fluctuations may be correlated with the inattentiveness and hyperactivity observed in ADHD patients \cite{bush2010attention}. Another significant finding of our study was the elevated temporal dynamics in the lateral parietal regions, including the parietal operculum. Moreover, the dysfunction in the parietal cortex was more widespread on the right side. Atypical dFC in the lateral parietal regions may be linked to an impaired capacity for top-down attention control in task-directed behaviour \cite{hale2014abnormal}. Additionally, right parietal abnormalities have been consistently reported in fMRI \cite{vance2007right}, EEG \cite{hale2014abnormal}, and neurogenetics \cite{hale2015parietal} studies, making them a potential biomarker for ADHD patients. Such rightward asymmetry of parietal dysfunction may be associated with a maturational deficit in children with ADHD at developmental stages \cite{vance2007right}.

In summary, we identified regions that showed higher temporal variability in functional connectivity in ADHD patients as compared to TD children. Our findings are consistent with the literature on the neural bases of ADHD and expand on the understanding of the disorder with a fresh perspective linking the behavioural inattention and hyperactivity observed in ADHD patients with the instability in functional connectivity in the brain.  
% It was observed that the FC among the nodes of the DMN showed high fluctuations signalling instability of introspective and resting states in ADHD patients. Atypical dFC was observed in the cingulo-temporal, fronto-parietal, and frontal attention networks. A rightward skew was observed in the parietal dysfunction 

\bibliographystyle{IEEEtran}
\bibliography{references.bib}

\end{document}